\documentclass{article}

\RequirePackage{amssymb}[1995/01/01]

\title{Quantum Field Symbolic Analog Computation: Relativity Model}
\author{A.~C.~Manoharan}

\begin{document}

\maketitle
\centerline{Turlock, CA}
\centerline{e-mail: Manoharan@worldnet.att.net}

\begin{abstract}

It is natural to consider a quantum system in the continuum limit of space-time
configuration.  Incorporating also, Einstein's special relativity, leads to
the quantum theory of fields.  Non-relativistic quantum mechanics and
classical mechanics are special cases. By studying vacuum expectation values
(Wightman functions $W(n; z)$ where $z$ denotes the set of $n$ complex
variables) of products of quantum field operators in a separable Hilbert
space, one is led to computation of holomorphy domains for these functions
over the space of several complex variables, ${\mathbb C}^n$.  
Quantum fields were reconstructed from these
functions by Wightman. Computer automation has been accomplished as
deterministic exact analog computation (computation over {\it cells} in the
continuum of ${\mathbb C}^n$) for obtaining primitive extended tube domains of
holomorphy.  This is done in a one dimensional space plus one dimensional time
model.
By considering boundary related
semi-algebraic sets, some analytic extensions of these domains are obtained by
non-deterministic methods.
The novel methods of computation raise interesting issues of
computability and complexity.  Moreover, the computation is independent of any
particular form of Lagrangian or dynamics, and is uniform in $n$, qualifying
for a universal quantum machine over ${\mathbb C}^\infty$.

\end{abstract}

\bibliographystyle{unsrt} 

\section{Introduction} 

Recently, there has been considerable interest in what is called {\em quantum
computation\/}~\cite{Feynman}.  The efforts in this regard are
to seek
improved ways of performing computations or building new types of computing
machines.  It is hoped that not only faster computation will be achieved~\cite
{Feynman, Shor}, but that better understanding of computational complexity will
come about~\cite{RMP, Vazirani}.

Quantum computers might appear to be discrete systems, such as a
finite collection of spins or qubits~\cite{Deutsch}.  But this is not the only
possibility.  Instead of thinking of a computer based on a physical system
which is understood in terms of non-relativistic quantum mechanics with spin
added on~\footnote{ Spin used to be added in a {\em ad hoc\/} fashion to
non-relativistic quantum mechanics until Dirac produced his relativistic
equation for the electron~\cite{Dirac}.}, one can also consider a system based
on relativistic quantum mechanics~\cite{mm}.

In another direction, the classical discrete digital computer~\cite{tu} has
been generalized to include the possibility of computing over the
continuum~\cite{sm2}.  This is because the latter way of computing over
the continuum is more appropriate to the way we do analysis, physics, and
engineering problems.  So this is a computing model which is based on
classical mechanics.  But classical mechanics could also be extended to
include relativity, getting relativistic mechanics~\cite{Einstein}.

Because, in studying atomic phenomena, classical mechanics has been replaced by
quantum mechanics, we could also think of more general models of
computing~\cite{mm, mmAMS} based on adding relativity to quantum theory to get
relativistic quantum field theory.

It is natural to consider our physical or quantum systems in the continuum
limit.  In fact Isaac Newton~\cite{Newton}, when studying gravitation, found
it natural to consider a continuous distribution of matter to model the earth's
gravitational action at external points.  From continuum quantum mechanics, by
combining relativity, we have quantum field theory.

\section{Generalized Quantum Computation}

The continuum limit (in space-time configuration) of quantum mechanics together
with the incorporation of Einstein's special relativity leads naturally to
the relativistic quantum theory of fields.  By taking the limit as the
velocity of light $c \rightarrow {\infty}$ we expect to get non-relativistic
quantum mechanics.  The limit as Planck's constant $h \rightarrow 0$ gives
classical mechanics~\cite{Jaffe}.  Quantum field theory includes all of
quantum mechanics, classical mechanics, and much more~\footnote{Examples
are, discrete anti-unitary symmetries, CPT invariance and the spin statistics
connection~\cite{sw}.}.  Included also will be unitary transformations and
superposition of amplitudes, which are regarded as prerequisites for quantum
computation.

It has been possible to generalize quantum computation to relativistic quantum
field computation in a certain model~\cite{ACMCombi}.  We expect other forms
of quantum computation, namely, those based on non-relativistic quantum
mechanics, topological quantum field theories~\cite{Freedman} and classical
mechanics~\cite{BCSS,Lloyd} to be related to this generalization.  The
relationship should shed light, not only on computing
possibilities~\cite{ACMtop, FKW}, but also on quantum field theory itself.

\section{Quantum Fields}

Non-relativistic quantum mechanics is not complete, because radiative
corrections have to be made to it, using field theory.  In dealing with a
system corresponding to an infinite number of degrees of freedom, it is well
known historically that formulations of quantum field theory like perturbation
theory lead to infinities resulting in the need for renormalization. 
Nevertheless, quantum electrodynamics has turned out to be, ``the most
accurate theory known to man"~\footnote{This statement is attributed to
Feynman.}.  Dirac, Schwinger and Feynman are some of the
principal contributers to quantum electrodynamics~\footnote{The spectacular
history of this is related in~\cite{SSS}} and hence to quantum field
theory~\cite{Weinberg}.  Relativistic covariance is of paramount importance in
correctly performing the renormalization process.

It is useful here to work within the Wightman formulation~\cite{Wightman,sw,jo}
of quantum field theory~\footnote{The fruitfulness and utility of this
formulation, from a current perspective, is available in~\cite{Kazhdan}}.
We are dealing with fields in the Heisenberg picture, without using
perturbation theory, nor any particular time frame related Hamiltonians.  The
theory is in terms of analytic functions (Wightman functions) of several
complex variables. These functions arise from their boundary values which are
vacuum expectation values of the form  $${\cal W}_m(x_1, x_2, \ldots x_m) = \\
 (\Omega ,\phi_1(x_1) \phi_2(x_2), \ldots \phi_m(x_m)\Omega)$$  of products
of $m$ quantum field operators in a separable Hilbert space. The field
operators transform according to appropriate unitary spin representations of
the Poincar\'e (inhomogenous $SL(2,{\mathbb C})$ group. Quantum fields are
uniquely reconstructed from these analytic functions by Wightman.

Let the ($m$-point) Wightman functions be denoted by $W(n; z)$ where $z$
denotes the set of $n$ complex variables. Here, $n = sm$ where $s \geq 2$ is
the space-time dimension; space-time will consist of 1-time and $(s-1)$-space
dimensions~\footnote {Because it is not known, at present, how to physically
understand concepts like closed time-like loops in more than one time
dimension, there will be only one dimension in time~\cite{Hawking}.}

Because these analytic functions are fundamental to the theory, one is led to
computation of holomorphy domains for these functions over the space of
several complex variables, ${\mathbb C}^n$~\cite{kw}.  The mass spectrum is
assumed to be reasonable, in the sense that momentum vectors $p^{\mu}$ lie in
the closed forward light cone, with time component $p^0 > 0$ except for the
unique vacuum state having $p = 0$.

\section{Computer Automation}
When $s=2$, i.e. in 1-dimensional space and 1-dimensional time, a system of
light-cone coordinates is appropriate~\cite{ru}.
In this 2-dimensional space-time, computer automation has been
accomplished as deterministic exact analog computation~\cite{ACMCombi}
(computation over ``cells" in the continuum of ${\mathbb C}^n$) to obtain
primitive extended tube domains of holomorphy for $W(n; z)$.
By a series of abstractions the computation is done with 
essentially reversible logic, programming in the Prolog language, and
simulating on a Turing machine.

Just as the classical computer, {\it Turing machine}, computes over ${\mathbb
Z}$ or equivalently over ${\mathbb Z}_2$, we now have what can be
called a {\it complex Turing machine}, in fact a, {\it severally
complex Turing machine}. 

The primitive extended tube domains are bounded by analytic hypersurfaces,
namely Riemann cuts denoted by $C_{ij}$ and other hypersurfaces denoted by
$S_{ij,kl}$ and $F_{ij,kl}$~\cite{ACMCombi}.  These domains are in the form of
semi-algebraic sets.  Since the computation is symbolic, it is also exact,
which is important in handling analytic functions.  

Because of Lorentz invariance properties of the physics involved, the domains
have a structure referred to as {\it complex Lorentz projective
spaces}~\footnote{This is different from Euclidean complex projective
spaces, well known in mathematics.  The difference is captured in the {\it
Hall-Wightman theorem}~\cite{sw}.}.  Related to this invariance are certain
continuum {\it cells} over which the computation occurs.  Thus this
computation is also like analog computation which would otherwise be regarded
as impossible to do exactly.

\section{Analytic Extensions}
In relativistic quantum field theory it is possible to implement the physical
requirement of microcausality.  There exists quantum microcausality (field
operators commute or anticommute) at totally space-like points.

Together with the requirement of permutation invariance of the domains, the
edge-of-the-wedge theorem provides enlargements of the original
primitive domains of analyticity to analyticity into unions of permuted
primitive domains.

Mapping these union domains creates some Boolean satisfiability
problems. In fact, the novel methods of computation raise interesting issues of
computability and complexity.
 
\section{Non-deterministic Holomorphic Extensions}
By the nature of analytic domains in more than one complex variable, it is in
general possible to further extend these domains towards the maximal domains
called {\it envelopes of holomorphy}.  By considering boundary related
semi-algebraic sets, there are non-deterministic computations of holomorphic
extensions of domains.  After the guessing step, the verification is by 
deterministic processes mentioned above.

Built-in permutation invariance 
has considerable power, 
just as $n!$ rapidly dominates over $2^n$ for large $n$.

\section{Uniformity of Computation}
Uniformity in the direction of universal computation has been discussed
in~\cite{BCSS}, in different contexts, including numerical analysis.  We
do have certain types of uniformity here.

First we note that the computation is independent of any particular form
of Lagrangian or dynamics, and is uniform in $n$, qualifying for a universal
quantum machine over ${\mathbb C}^\infty$.  The latter space is the infinite
discrete union $\bigsqcup^\infty_{n = 1}{\mathbb C}^n$.

\subsection{Function Order Uniformity}
When the progrm runs for $s=2$, dynamic memory allocation is used through the
operating system.  Because $n$ can be input as a variable, only part of the
whole memory management cost is outside the program.  The program itself is
independent of $n = sm$ and therefore is uniform in $n$, which is unbounded
above.  We can call this {\it function order uniformity in} $n_\infty$.

\subsection{Space-time Uniformity}

In addition, there is uniformity in the dimension $s \geq 2$ of space-time, in
the following manner.
Given a dimension $s \ge 2$ of space-time, looking at the
semi-algebraic sets defining the primitive extended tube domains of
holomorphy (hypersurface boundaries), and at function orders, there are three
different classes of orders.  These classes comprise, a) lower order W
functions,  b) intermediate order W functions, and c) high order W
functions~\cite{acmcolorado}.  Extended tube domains for all high order W
functions have the same complicacy.  For a)
we have $m \le s + 1$, and for c), $m > s(s-1)/2 + 2$.  The remaining
cases lie in class b).  For example, there is no class b) for $s = 2$, the most
complicated primitive domain being for the 3-point function.  If $s = 3$, then
$m = 5$ is the only case in class b).  When $s = 4$, we have in class b), the
cases, $m = 6, 7$ and $8$.

Since $s \ge 2$ is unbounded above, we can call this {\it space-time dimension
uniformity in} $s_\infty$.

\subsection{co-NHolo Uniformity}

The holomorphy envelopes $H[D_m]$ for different orders $m$ of Wightman
functions are related~\cite {acm} in the following way.

{\it For} $0 < r < m$, and {\it relative to} ${ H[D_m] }$,
$${ H[D_m] \subset \\ 
 \bigcap_{\sigma \in {\rm P}_m} \\  
  \left\{ H [\sigma D_{m-r}] \times 
  (\sigma {\mathbb C}^{sr})  \right\}  },$$ 

where $\sigma$ denotes permutations in ${\rm P}_m$, the permutation group in
the $m$ points of the $m$-point W function.  In the case of {\it Schlicht}
domains (analogous to single sheeted Riemann surfaces in $\mathbb C$), the
$\subset$ sign means set theoretic inclusion.

For example, in $s=2$, the 4-point function cannot be continued beyond the
2-point function Riemann cuts nor the (permuted) 3-point function
K\"all\'en-Wightman domains of holomorphy.

This is a statement regarding analyticity that does not exist, and thus refers
to the complements of the domains of holomorphy;  hence the use of the prefix
{\it co-}.  Because computations of analytic extensions of domains are
non-deterministic (hence the notation {\em N}), we can say that we have {\it
co-NHolo} uniformity.  

\section{Conclusion}
We started with relativity and continuum quantum field theory.  Arbitrary
numbers of particles (optionally with spins) can be created or annihilated. 
Relying on a fruitful set of models, we have related what appeared to be
different models of quantum and classical computation based on
non-relativistic quantum mechanics and classical mechanics.  Exact
deterministic and non-deterministic computation over continuous domains appear
naturally.  Furthermore there is uniformity in computation over, unbounded
above, or arbitrarily high order $n$ of $W(n;z)$ and arbitrarily high
dimension $s$ of space-time.  In the present context this can be called {\em
co-NHolo} uniformity in $n_\infty$ and $s_\infty$.  The novel methods of
computation raise interesting issues of computability and complexity, and
possibly could shed more light on quantum field theory itself.

\end{document}